\documentclass[superscriptaddress, twocolumn, prb]{revtex4-2}
\usepackage{graphicx}
\usepackage{here}

\newcommand{\NIMS}{National Institute for Materials Science, 1-2-1 Sengen, Tsukuba, Japan}
\newcommand{\UT}{University of Tsukuba, 1-1-1 Tennoidai, Tsukuba, Japan}

\begin{document}
\title{Experimental exploration of ErB$_2$ and SHAP analysis on a machine-learned model of magnetocaloric materials for materials design}

\author{Kensei Terashima}
\email{TERASHIMA.Kensei@nims.go.jp}
\affiliation{\NIMS}
\author{Pedro Baptista de Castro}
\affiliation{\NIMS}
\affiliation{\UT}
\author{Akiko T Saito}
\affiliation{\NIMS}
\author{Takafumi D Yamamoto}
\email{Present address: Tokyo University of Science, Tokyo, Japan}
\affiliation{\NIMS}
\author{Ryo Matsumoto}
\affiliation{\NIMS}
\author{Hiroyuki Takeya}
\affiliation{\NIMS}
\author{Yoshihiko Takano}
\affiliation{\NIMS}
\affiliation{\UT}


\begin{abstract}
  Stimulated by a recent report of a giant magnetocaloric effect in HoB$_2$ found via machine-learning predictions, we have explored the magnetocaloric properties of a related compound ErB$_2$, that has remained the last ferromagnetic material among the rare-earth diboride (REB$_2$) family with unreported magnetic entropy change $|\Delta S_M|$. The evaluated $|\Delta S_M|$ at field change of 5 T in ErB$_2$ turned out to be as high as 26.1 (J kg$^{-1}$ K$^{-1}$) around the ferromagnetic transition (${T_C}$) of 14 K. In this series, HoB$_2$ is found to be the material with the largest $|\Delta S_M|$ as the model predicted, while the predicted values showed a deviation with a systematic error compared to the experimental values. Through a coalition analysis using SHAP, we explore how this rare-earth dependence and the deviation in the prediction are deduced in the model. We further discuss how SHAP analysis can be useful in clarifying favorable combinations of constituent atoms through the machine-learned model with compositional descriptors. This analysis helps us to perform materials design with aid of machine learning of materials data.
\end{abstract}
\maketitle

\section{Introduction}
The arrangement of spin in solids and molecules provides a source of an entropy change upon application or removal of the applied field, namely the magnetocaloric effect. Such an effect tends to peak around the magnetic ordering temperature of the material, therefore the magnitude of the effect is known to be highly temperature- and material-dependent \cite{Gschneidner2000, Franco2018}.  One of common measures for this effect is the magnitude of magnetic entropy change caused by a change in the external field to the material ($|\Delta S_M|$), and currently, there is an increasing demand for high $|\Delta S_M|$ materials as it would provide an alternative cooling route other than conventional gas \cite{Franco2018}. The refrigeration using the magnetocaloric effect has been argued to be highly suitable for low-temperature applications \cite{Numazawa}, such as the liquefaction of hydrogen. This in turn, suggests the particular importance of materials with high $|\Delta S_M|$ around hydrogen liquefaction temperature ($\sim$20.3 K). \\

Recently, we have developed a machine-learned model \cite{Castro2020-12} that relates compositions of materials to their peak values of $|\Delta S_M|$ ($|\Delta S_M|^{peak}$) with a mean absolute error of 1.8 (J kg$^{-1}$ K$^{-1}$) by using reported $|\Delta S_M|^{peak}$ data \cite{Franco2018}. In the model, compositional descriptors \cite{Ward1, XenonPy} generated by XenonPy package \cite{XenonPy} were used to train the model, where corresponding descriptors for each composition are made by applying 7 kinds of mathematical operations (arithmetic mean, geometrical mean, harmonic mean, variance, sum, max, min, weighted by compositions) to 58 physical properties of atomic elements, in addition to the counting of constituent atoms and experimental values of appied field change $\mu_0 \Delta H$. We used XGBoost package \cite{XGBoost} to build a model with a gradient boosting-based decision tree algorithm. Once a model is built, such a compositional descriptors-based model can be readily applied for the prediction of target materials in the search space as it requires only compositions to predict with fixed value of $\mu_0 \Delta H$ (5 T). Therefore, the constructed model was applied to materials with unknown $|\Delta S_M|^{peak}$ values for the selection of a candidate to be examined, leading to an experimental discovery of a gigantic magnetocaloric effect in HoB$_2$ with $|\Delta S_M|^{peak}$ reaching 40.1 (J kg$^{-1}$ K$^{-1}$) around \textit{T} = 15 K under field change of 5 T \cite{Castro2020-12}. In addition, it was also found that HoB$_2$ undergoes an additional magnetic transition at \textit{T} = 11 K, which also contributes to $|\Delta S_M|$.\\

HoB$_2$ is one of the series of rare-earth diborides REB$_2$ with hexagonal AlB$_2$ type structure (RE = Tb, Dy, Ho, Er, Tm, Yb, Lu) \cite{Buschow1977}. Except for RE = Yb \cite{Avila_2003} and Lu, they are known to exhibit ferromagnetic ordering($T_C$ = 151, 55, 15, 16, 7 K for RE = Tb, Dy, Ho, Er, Tm, respectively). Several of their detailed magnetic properties and magnetocaloric effect have been further reported. Among them, TbB$_2$ \cite{Han_2010} and TmB$_2$ \cite{Mori2009} show single ferromagnetic transition, while DyB$_2$ \cite{Meng2012} and HoB$_2$ \cite{Castro2020-12} show an additional magnetic transition below $T_C$, where for the case of HoB$_2$ it has been clarified to be related with spin reorientation \cite{Terada2020}. The magnitude of $|\Delta$S$_M|^{peak}$ (J kg$^{-1}$ K$^{-1}$) at 5 T has been reported to be 10, 17, 40 for RE = Tb \cite{Han_2010}, Dy \cite{Meng2012}, Ho \cite{Castro2020-12}, and 24 for Tm \cite{Mori2009} (estimated from fitting a power law curve \cite{Franco_scaling2008} on the data calculated by using the specific heat data reported at \cite{Mori2009}).  Contrary to the research above, the detail of magnetic and magnetocaloric properties of ErB$_2$ has been left vailed.  Therefore, a comparative study with ErB$_2$ would help a comprehensive understanding of REB$_2$ series that hosts a giant magnetocaloric effect.

In this paper, we report the magnetic properties and magnetocaloric effect of ErB$_2$.  The evaluated $|\Delta S_M|^{peak}$ was as high as 26.1 (J kg$^{-1}$ K$^{-1}$), which is lower than that of HoB$_2$ (40.1 (J kg$^{-1}$ K$^{-1}$)).  The RE dependence in the magnitude of the magnetocaloric effect in REB$_2$ has been shown to peak at RE = Ho, similarly to the model. However, a lack of quantitative agreement was also found between the prediction and the experimental results.  Through a model analysis where coalitions to the prediction are resolved for each descriptor, we discuss how the machine-learned model tried to address the atomic-species dependence of magnetocaloric effect such as RE atoms.  We also discuss that such a coalition analysis for compositional descriptors in machine-learned models would help researchers to identify favorable atoms for enhancing target property of materials with the aid of constructed models.\\

\begin{figure}[h]
  \centering
  \includegraphics[width=7 cm]{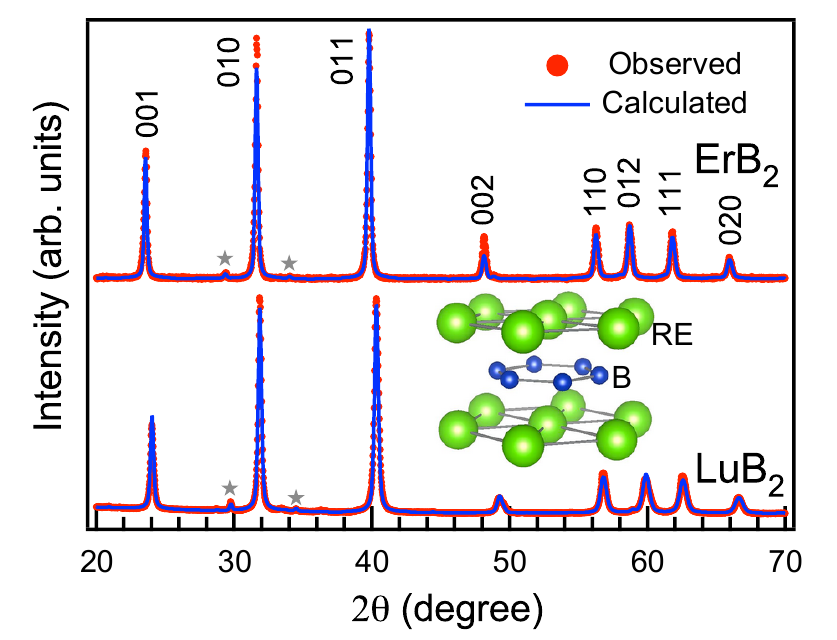}
  \caption{XRD pattern of ErB$_2$ (top) and LuB$_2$ (bottom) samples. Asterisks in the figure denote Er$_2$O$_3$ or Lu$_2$O$_3$ impurity. Dots correspond to the experimental data, while lines correspond to the calculated pattern by Rietveld analysis. Inset: Crystal structure of REB$_2$, drawn by VESTA software \cite{VESTA}}
  \label{fig1}
\end{figure}

\section{Methods}

Polycrystalline samples of ErB$_2$ and LuB$_2$ were obtained by arc melting of Er (99.9$\%$) or Lu (99.9$\%$) and B (99.9$\%$) in an evacuated arc furnace with Ar atmosphere.  Samples were melted several times to ensure homogeneity. X-ray diffraction (XRD) patterns of samples were measured using a MiniFlex600 (Rigaku) and analyzed by using FullProf package \cite{FullProf}.  Magnetization of ErB$_2$ sample was obtained by using MPMS (Quantum Design), while specific heat of ErB$_2$ and LuB$_2$ were obtained using PPMS (Quantum Design) apparatus.

\section{Results and Discussions}

\subsection{Physical properties of ErB$_2$}\label{expt}

\begin{figure}[h]
  \centering
  \includegraphics[width=7 cm]{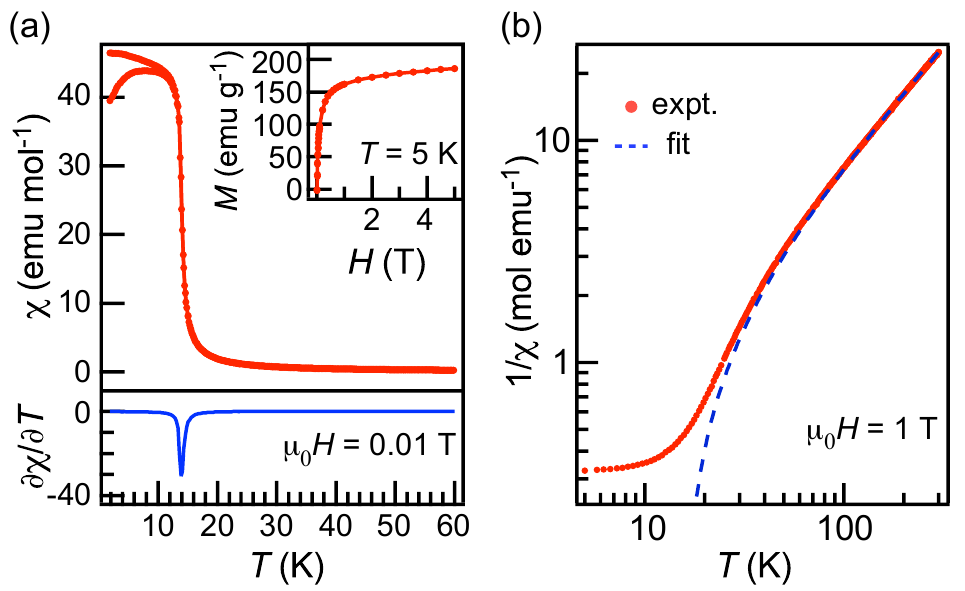}\hspace{5pt}
  \caption{(a) Temperature dependence of magnetic susceptibility in ErB$_2$ (top) and its temperature-derivative (bottom), taken at $\mu$$_0$$H$ = 0.01 T. Inset shows isothermal magnetization curve of ErB$_2$, taken at $T$ = 5 K. (b) Inverse magnetic susceptibility of ErB$_2$ at 1 T. The dashed line in the figure is the fit to Curie-Weiss law in the temperature range of 200-300 K.}
  \label{fig2}
\end{figure}

Figure 1 shows the XRD patterns of obtained samples. In addition to the target material ErB$_2$ (top of Figure 1), we also fabricated a nonmagnetic reference LuB$_2$ (bottom of Figure 1).  From the Rietveld analyses, the obtained samples have been evaluated to hold 99 \% main phase of ErB$_2$ and 1 \% of Er$_2$O$_3$ for ErB$_2$, and 98 \% main phase of LuB$_2$ and 2 \% of Lu$_2$O$_3$ for LuB$_2$.  The obtained lattice constants were 3.2725(2) $\mathrm{\mathring{A}}$ (a-axis) and 3.7845(3) $\mathrm{\mathring{A}}$ (c-axis) for ErB$_2$ and 3.2387(1) $\mathrm{\mathring{A}}$ (a-axis) and 3.6941(1) $\mathrm{\mathring{A}}$ (c-axis) for LuB$_2$, showing an overall agreement with literature \cite{Buschow1977}.\\

\begin{figure*}[t]
  \centering
  \includegraphics[width=14.5 cm]{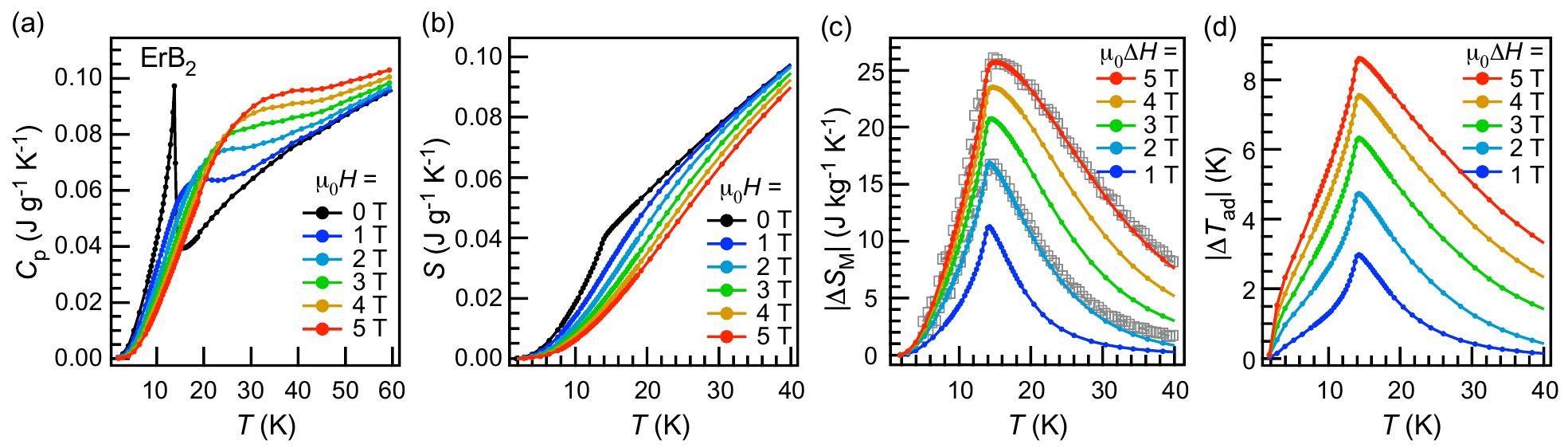}\hspace{5pt}
  \caption{(a) Specific heat of ErB$_2$ under various magnetic fields. (b) Entropy curves of ErB$_2$ under magnetic fields deduced from data in (a). (c) $|\Delta S_M|$ of ErB$_2$, obtained from data in (b). Gray open squares denote $|\Delta S_M|$ estimated from magnetization curves (supplemental information S1).  (d) $\Delta T_{ad}$ of ErB$_2$, obtained from data in (b).}
  \label{fig3}
\end{figure*}

In Figure 2(a), we show the isofield magnetic susceptibility as a function of temperature ($\chi-T$ curve) for both zero field cooling (ZFC) and field cooling (FC) on ErB$_2$, taken under a magnetic field of 0.01 T.  The bottom of Figure 2 shows $\partial$$\chi$/$\partial$$T$ of field-cooled $\chi-T$ curve, indicating that ErB$_2$ shows a single ferromagnetic transition at Curie Temperature (\textit {$T_C$})$\sim$14 K with an absence of lower transition as observed in DyB$_2$ and HoB$_2$ down to 1.8 K.  As in the inset, isothermal magnetization as a function of field ($M-H$ curve) at $T$ = 5 K on ErB$_2$ did not show a clear hysteresis.  Not only the $M-H$ curve, but there is also only a small difference between ZFC and FC $\chi-T$ curves below $T_C$ that might be coming from a domain effect.  Figure 2(b) shows 1/$\chi$ as a function of temperature up to room temperature, taken at 1 T and shown in a logarithmic scale. The dashed line in the figure corresponds to the result of fitting the data by Curie-Weiss plot between the temperature range from 200 to 300 K.  The estimated Weiss temperature $\Theta$ was 15.4 K, in good agreement with Curie temperature estimated from the dip in the $\partial$$\chi$/$\partial$$T$ curve.  The deviation between the fitting curve and the experimental data becomes evident below $\sim$ 80 K, which may be attributed to the effect of the crystal electric field \cite{Mori2009}. The estimated effective magnetic moment $\mu_{eff}$ was 9.57 $\mu_B$ per formula unit, which is close to that of the theoretical value for the free trivalent Er$^{3+}$ atom (9.59).

Next, we discuss the magnetocaloric effect of ErB$_2$. Figure 3(a) shows the temperature dependence of specific heat on ErB$_2$ taken at various fields, in the temperature range between 1.8 and 60 K. Under zero field, ErB$_2$ exhibits a peak in specific heat around 14 K that corresponds to the ferromagnetic transition, and the peak gets smooth and its position migrates to higher temperatures as we apply a magnetic field.  This behavior is of typical ferromagnets.  Entropy $S$ curves estimated from the specific heat results are shown in Figure 3(b).  In such $S-T$ curves, the vertical gap between two curves under different magnetic fields corresponds to $|\Delta S_M|$, while the horizontal gap between two curves corresponds to the magnitude of adiabatic temperature change (so-called $|\Delta T_{ad}|$), that are shown in Figure 3(c) and 3(d), respectively.  $|\Delta S_M|$ can be also estimated from a series of magnetization measurements (see supplemental information S1 for detail), and that for 2 T and 5 T are in Figure 3(c) as gray open squares, showing that the $|\Delta S_M|$s estimated from two experimental methods agree very well. The $|\Delta S_M|$$^{peak}$ of this material is estimated to be 26.1 (J kg$^{-1}$ K$^{-1}$) at $T$$\sim$14 K. In volumetric unit, it corresponds to 0.23 (J cm$^{-3}$ K$^{-1}$) as the density of the sample was evaluated to be 8.937 (g/cm$^3$) by the XRD analysis.  The overall shape of the temperature dependence of $|\Delta S_M|$ among different applied fields is almost unchanged, which is a typical feature of second-order ferromagnetic materials \cite{Franco_scaling2008}.  Also, the peak value of $\Delta T_{ad}$ of ErB$_2$ turned out to be 8.6 K at $T$$\sim$14 K under field change of 5 T.

\begin{figure}[h]
  \centering
  \includegraphics[width=7 cm]{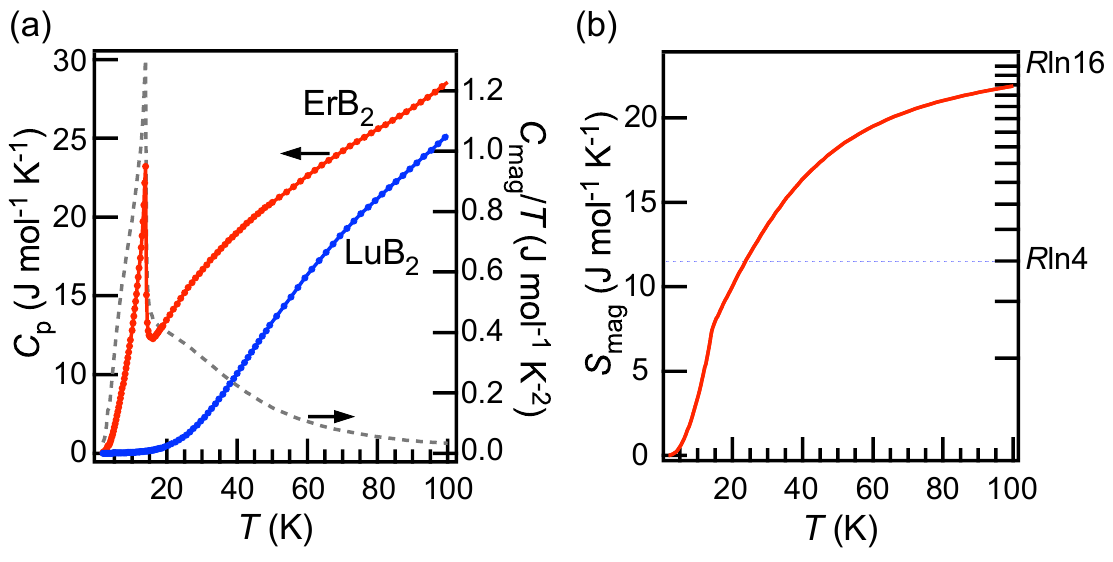}\hspace{5pt}
  \caption{(a) Specific heat of ErB$_2$ and LuB$_2$ (left axis, solid lines) and estimated C$_{mag}$/$T$ (right axis, dashed line). (b) Magnetic entropy curve of ErB$_2$ as a function of temperature.}
  \label{fig4}
\end{figure}

Figure 4(a) shows the comparison of specific heat between ErB$_2$ and nonmagnetic reference LuB$_2$ in a wider temperature range.  Apart from a peak around 14 K due to ferromagnetic transition, the specific heat of ErB$_2$ shows broad bumps or kink structures around 40-60 K. This is the temperature range that corresponds well to where there is a deviation from the Curie-Weiss fit in the 1/$\chi$ curve, implying that these are due to the presence of energy levels split caused by crystal field effect.  By integrating the difference in specific heat between ErB$_2$ and LuB$_2$ divided by temperature ($C_{mag}/T$, shown as the dashed line in Figure 4(a)) over temperature, we have estimated the temperature dependence of magnetic entropy $S_{mag}$ of ErB$_2$ as shown in Figure 4(b).  The estimated $S_{mag}$ at $T_C$ is 7.6 J/mol, that corresponds to 65 $\%$ of $R\ln$4, implying that most likely 2 degenerate Kramers doublet levels of Er$^{3+}$ are involved for the observed magnetic entropy change.

\begin{figure}[h]
  \centering
  \includegraphics[width=7 cm]{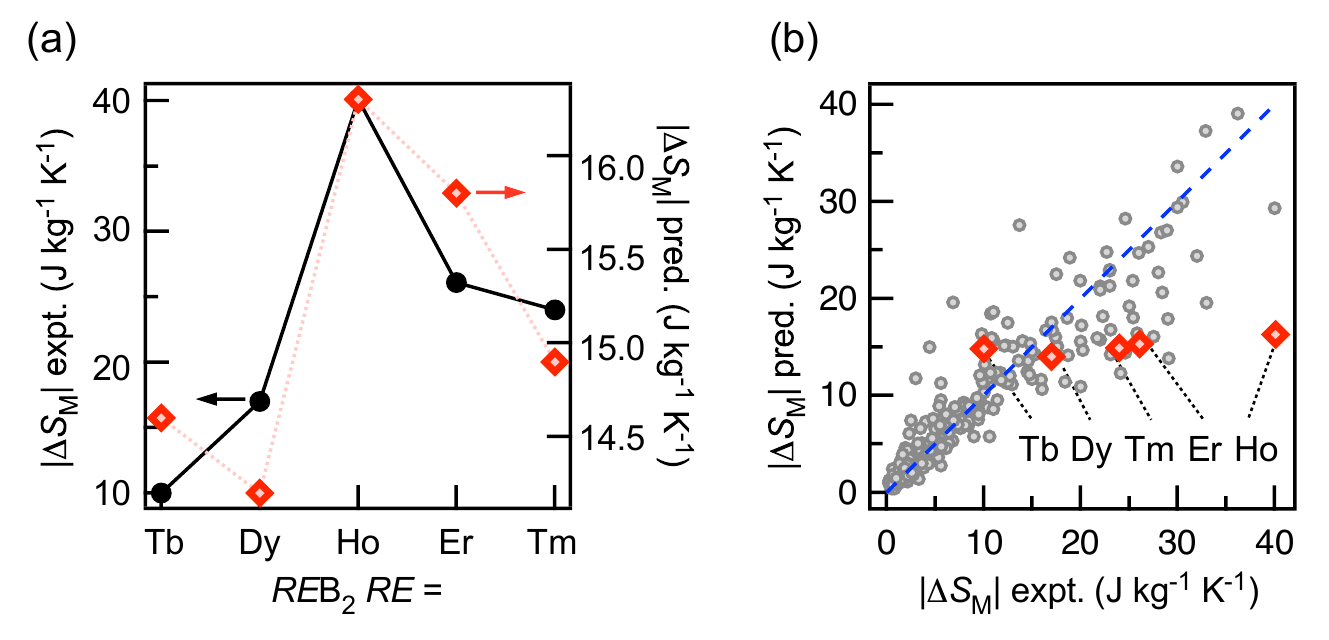}\hspace{5pt}
  \caption{(a) Comparison of $|\Delta S_M|^{peak}$ values in REB$_2$ (RE = Tb, Dy, Ho, Er, Tm) at $\mu_0 \Delta H$ = 5 T between experiment (left axis, black filled circles, \cite{Han_2010, Meng2012,Castro2020-12,Mori2009} and this work) and prediction by the machine-learned model (right axis, red open squares). (b) $|\Delta S_M|^{peak}$ values predicted by machine-learned model plotted against the experimental values for REB$_2$ (red open squares) and those in the test dataset for the model (gray circles) \cite{Castro2020-12}.}
  \label{fig5}
\end{figure}

\subsection{SHAP analyses of machine-learned model}

Here let us compare the rare-earth dependence of $|\Delta S_M|^{peak}$ in REB$_2$ system at field change of 5 T between experiment and prediction by the machine-learned model \cite{Castro2020-12}, where all REB$_2$ are not in the training dataset, \textit{i.e.} the entire series are unseen to the model.  As we have clarified in Section \ref{expt} and summarized in Figure 5(a) (black filled circles, left axis), $|\Delta S_M|^{peak}$ in stoichiometric REB$_2$ is maximized at RE = Ho case.  On the other hand, the model (red open squares, right axis) also predicts that RE = Ho would show the highest $|\Delta S_M|^{peak}$ despite the scale of values being quite different from the experimental values.  The same trend is also visible in Figure 5(b) where we plot $|\Delta S_M|^{peak}$ of REB$_2$ predicted by the model against the experimental values, with those in the test dataset \cite{Castro2020-12}. In the figure, the difference between the prediction and experimental values tends to be greater as $|\Delta S_M|^{peak}$ gets higher, indicating that the model has a certain systematic error in the prediction for REB$_2$ system. The lack of quantitative accuracy in the prediction of the model may indicate that there could be a novel mechanism for enhancing the magnetocaloric effect in this series, that was not in the materials of the training dataset of the model.  Now it would be intriguing to ask the model how it predicted this RE-dependence, in other words, what compositional descriptors made the model predict that HoB$_2$ is the most hopeful in REB$_2$, but failed to predict with quantitative accuracy. From now on, we will show that in addition to the conventional dependence plot between the target and the compositional descriptors, a dependence plot between the coalition of each descriptor to the individual prediction by the model and a visualization of the elemental property library would help us to understand the underlying insight that the model has captured. We show two use cases as an example for this approach: first, to understand the local explanation of RE-dependence in the prediction of REB$_2$; second, to understand the global tendency to achieve the high target value.

\begin{figure*}[!htpb]
  \centering
  \includegraphics[width=14.0 cm]{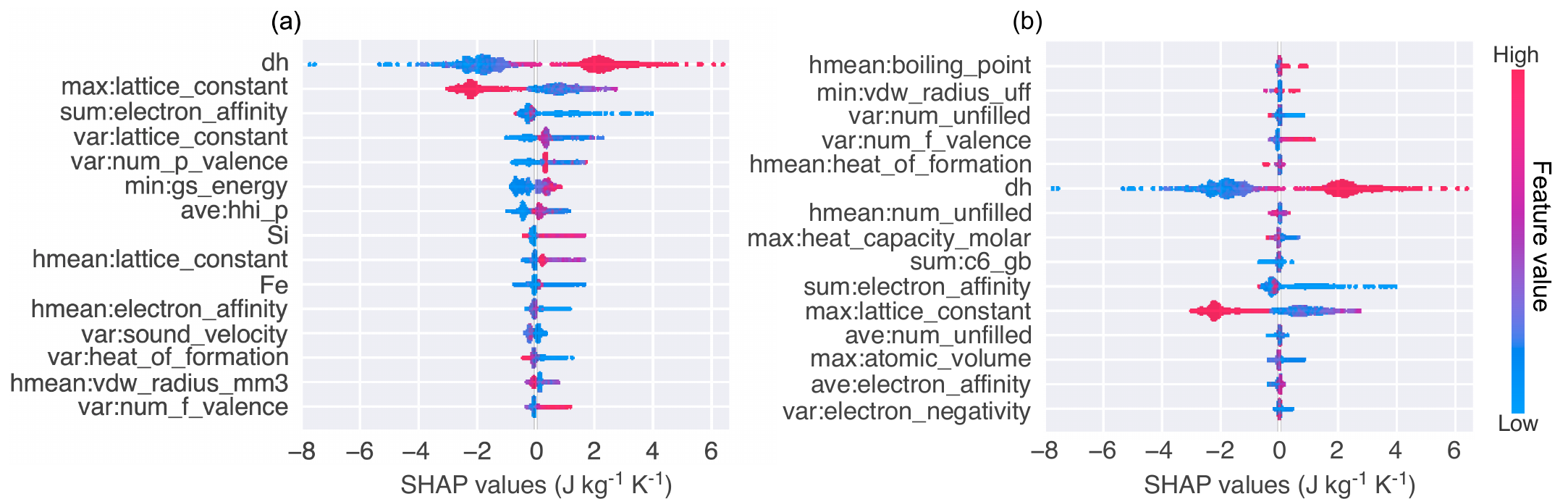}
  \caption{Summary plots of SHAP values up to the top 15 compositional descriptors, sorted by (a) magnitude of the absolute mean (i.e. significance), and (b) magnitude of RE-dependence in REB$_2$ prediction. Dots in the figure represents each data point of the training dataset.}
  \label{fig6}
\end{figure*}

Such coalition values are accessible by calculating the Shapley values from the constructed model \cite{NIPS2017_7062}. For this purpose, we applied a SHapley Additive exPlanations (SHAP) analysis \cite{NIPS2017_7062} to the model by using the TreeSHAP package \cite{lundberg2020local2global}.  The calculated SHAP value $\Phi_{i,j}$ for $i$-th descriptor of $j$-th data point represents the amount of the shift in the predicted target value for $j$-th data compared with the mean target value of the entire training data ($\Phi_{mean}$, in our case 7.48 J kg$^{-1}$ K$^{-1}$), by disclosing $i$-th descriptor value to the model when all other descriptor values has been informed to the model. One of the prominent characteristics of SHAP values is that they are additive, namely the final predicted value $\Phi_j$ for data point $j$ can be derived by simply summing up $\Phi_{i,j}$ for all descriptors $i$:\\
\begin{center}
  $\Phi_j$ = $\Sigma_i \Phi_{i,j}$ + $\Phi_{mean}$\\
\end{center}

Therefore, we can resolve what descriptor $i'$ made a significant contribution to the prediction of the model for individual entry $j'$.  The analysis can be also applied to the training dataset, to depict how the model understood the data.  Figure 6 shows summary plots of SHAP values for the training dataset, where compositional descriptors are sorted by (a) the magnitude of absolute mean and (b) the magnitude of RE-dependence (RE = Tb, Dy, Ho, Er, Tm) in prediction for REB$_2$ series, and shown up to top 15 for each.  The latter is obtained by sorting the SHAP values of descriptors by the magnitude of the standard deviation in the prediction of REB$_2$ compounds.  In Figure 6(a), the biggest contribution to the predicted values came from ``dh" which stands for $\mu_0 \Delta H$ namely the magnitude of field change.  For this ``dh", a high descriptor value tends to give high SHAP values, which makes sense as $|\Delta S_M|$ generally becomes higher by increasing the magnitude of the operating field change, especially in case of second-order magnetic materials \cite{Franco_scaling2008}. On the other hand, for the rest of the descriptors, SHAP values show complicated dependencies on descriptors as we will discuss some of them in detail later.

\begin{figure}[h]
  \centering
  \includegraphics[width=8.6 cm]{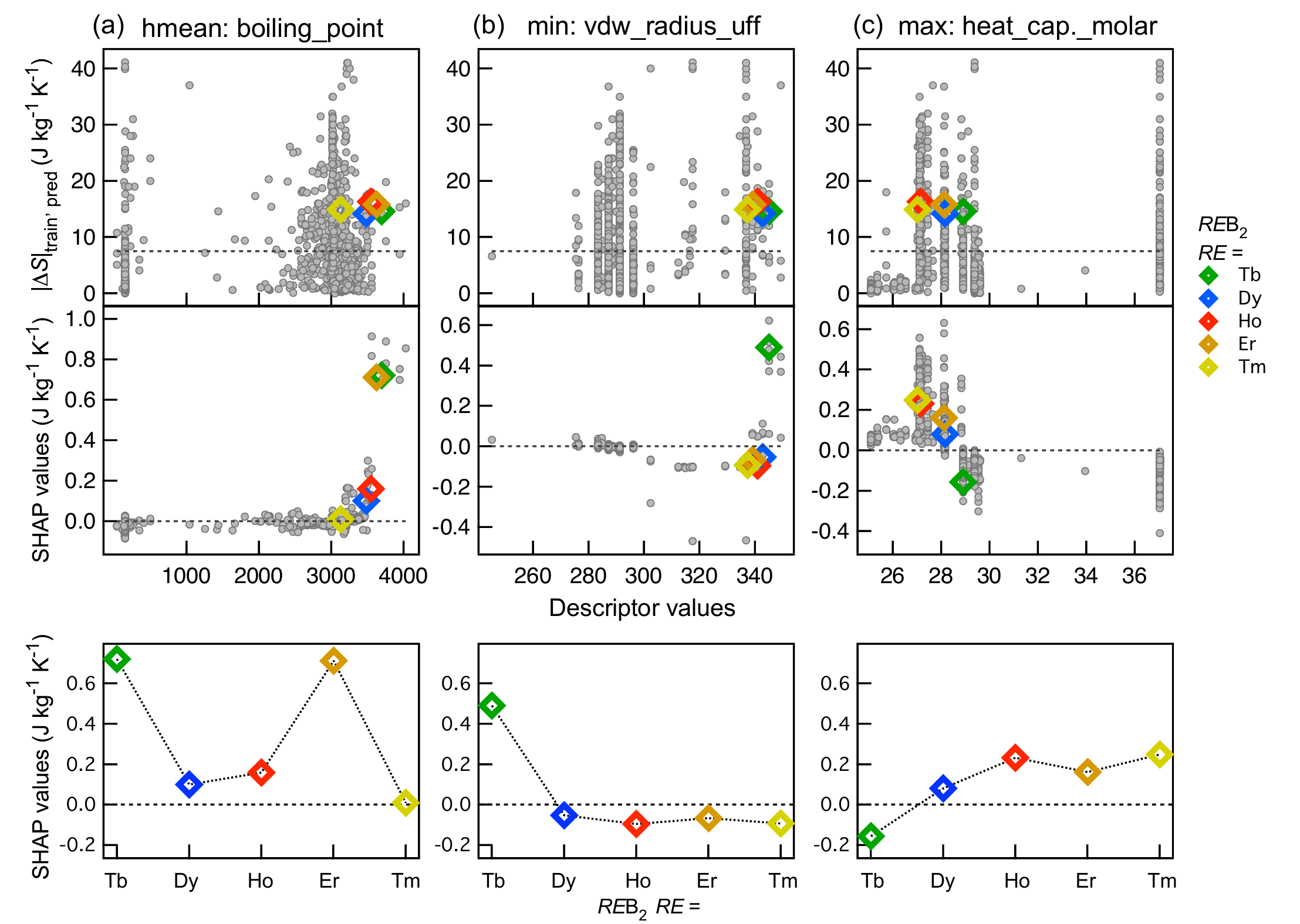}\hspace{5pt}
  \caption{Top panel: Target values in training data (gray circles) plotted against focused compositional descriptors of (a) harmonic mean of boiling point, (b) minimum of van der Waals radius, and (c) maximum of molar heat capacity. Horizontal dashed lines correspond to the mean target value of training data (7.48). Predicted values for REB$_2$ are also shown as colored open squares. Middle panel: The same for SHAP values for corresponding compositional descriptors. Bottom panel: RE-dependence of SHAP values for corresponding compositional descriptors in the prediction for REB$_2$}
  \label{fig7}
\end{figure}

Interestingly, descriptors that give high RE-dependence for the prediction in REB$_2$ (Figure 6(b)) do not necessarily possess high SHAP values (Figure 6(a)).  Here we will discuss more details about the origin of RE-dependence in the model. Figure 7 shows the comparison of target values in the train data and SHAP values, together with RE-dependence of prediction in REB$_2$ series for three representative descriptors, namely harmonic mean of boiling point (K), minimum of van der Waals radius (pm), and maximum of molar heat capacity (J K$^{-1}$ mol$^{-1}$) of constituent atoms. Those three descriptors were chosen from the list in Figure 6(b) so that they have different atomic species dependence, to show here examples for various adaptivity of descriptors.  Dashed lines in the top panel are the mean value of the target in the training dataset (7.48 J kg$^{-1}$ K$^{-1}$). Therefore, the SHAP values in the middle panel correspond to the degree of the shift from this mean value in the model. The vertical dispersion in SHAP plots comes from SHAP interaction between descriptors \cite{lundberg2020local2global} (A specific example of interaction term is shown in supplemental information S2).  The top panel figures can be drawn even before the construction of the machine-learned model, but it is quite difficult to figure out only from such plots how these descriptor values are related to the target property. Predicted values for REB$_2$ are overlaid in the figure, though it is also hard to draw any conclusion regarding how RE-dependence is predicted in these plots.

\begin{figure}[h]
  \centering
  \includegraphics[width=7 cm]{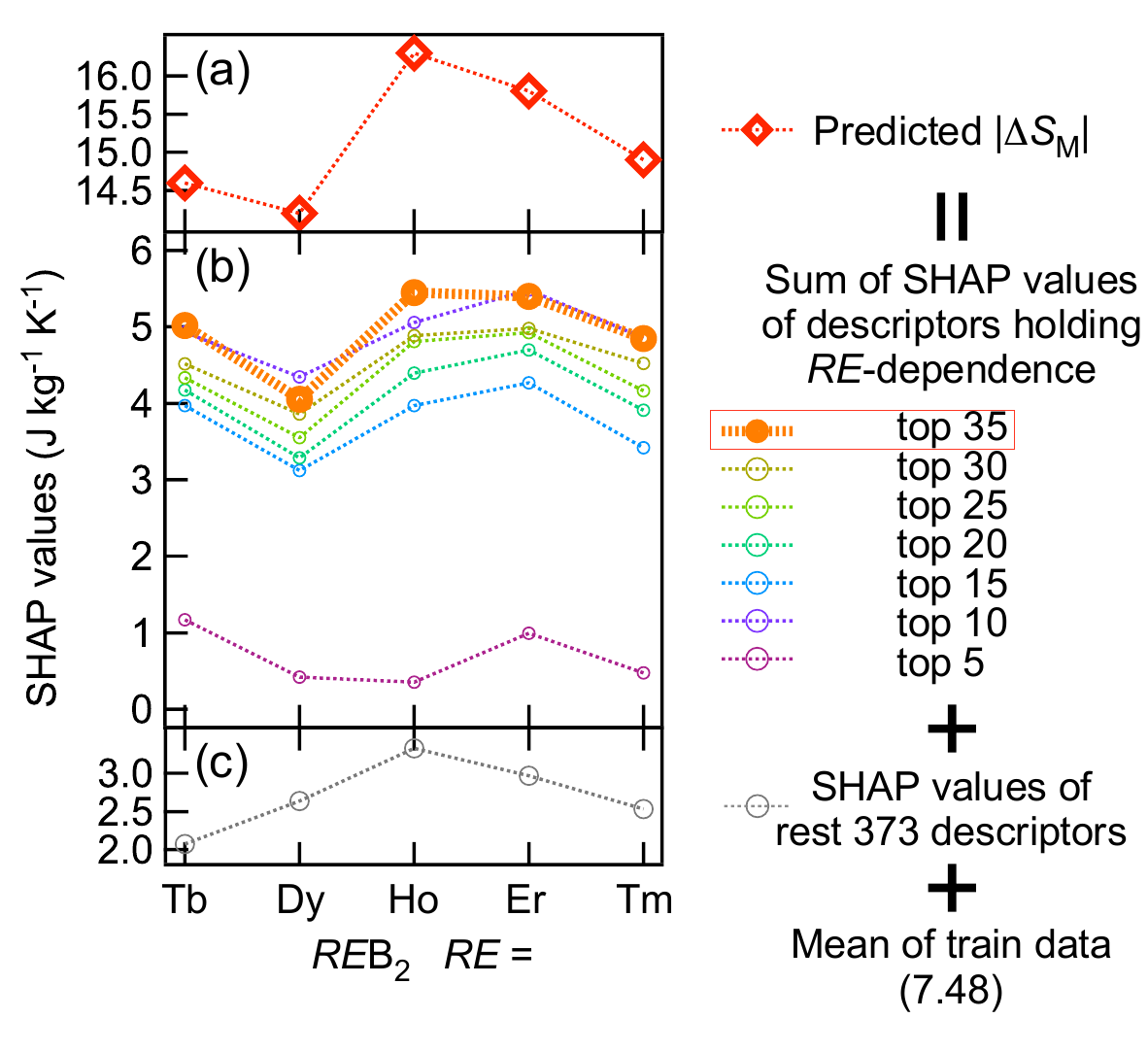}\hspace{5pt}
  \caption{RE-dependence of SHAP values. (a) Total sum of SHAP values that is identical to predicted target values. (b) Partial sum of SHAP values. (c) Sum of SHAP values for the rest of descriptors.}
  \label{fig8}
\end{figure}

On the other hand, as in the middle panel, the relationship between descriptors and the partial predicted value associated with each descriptor becomes clearly visible through SHAP analysis.  In the bottom panel, we show the RE-dependence of the SHAP values for each descriptor.  Among those descriptors, harmonic mean of boiling point showed the largest RE-dependence.  However, the RE-dependence of the SHAP value for boiling point does not coincide with that of the final prediction, and neither do other descriptors as well.  Figure 8 shows predicted target values for REB$_2$, resolved in partial sums of the SHAP values for descriptors.  As shown in Figure 8, until we sum up the SHAP values for the top 35 descriptors sorted by RE-dependence in the SHAP values, the predicted value for HoB$_2$ do not become the largest among REB$_2$.  Therefore, it can be said that the RE-dependence in the predicted values is based on a number of descriptors, and it is difficult to mention which is a key descriptor that made HoB$_2$ to be the most promising in the prediction, at least in the current model. It is also possible that the suppressed amplitude of the prediction in REB$_2$ series might be a result of the sum of small contributions of SHAP values with different RE-dependence (even though it is a consequence of learning of the training dataset), which ended up in lack of quantitative accuracy for the prediction of $|\Delta S_M|$ in this series.

\begin{figure*}[!htpb]
  \centering
  \includegraphics[width=14.5 cm]{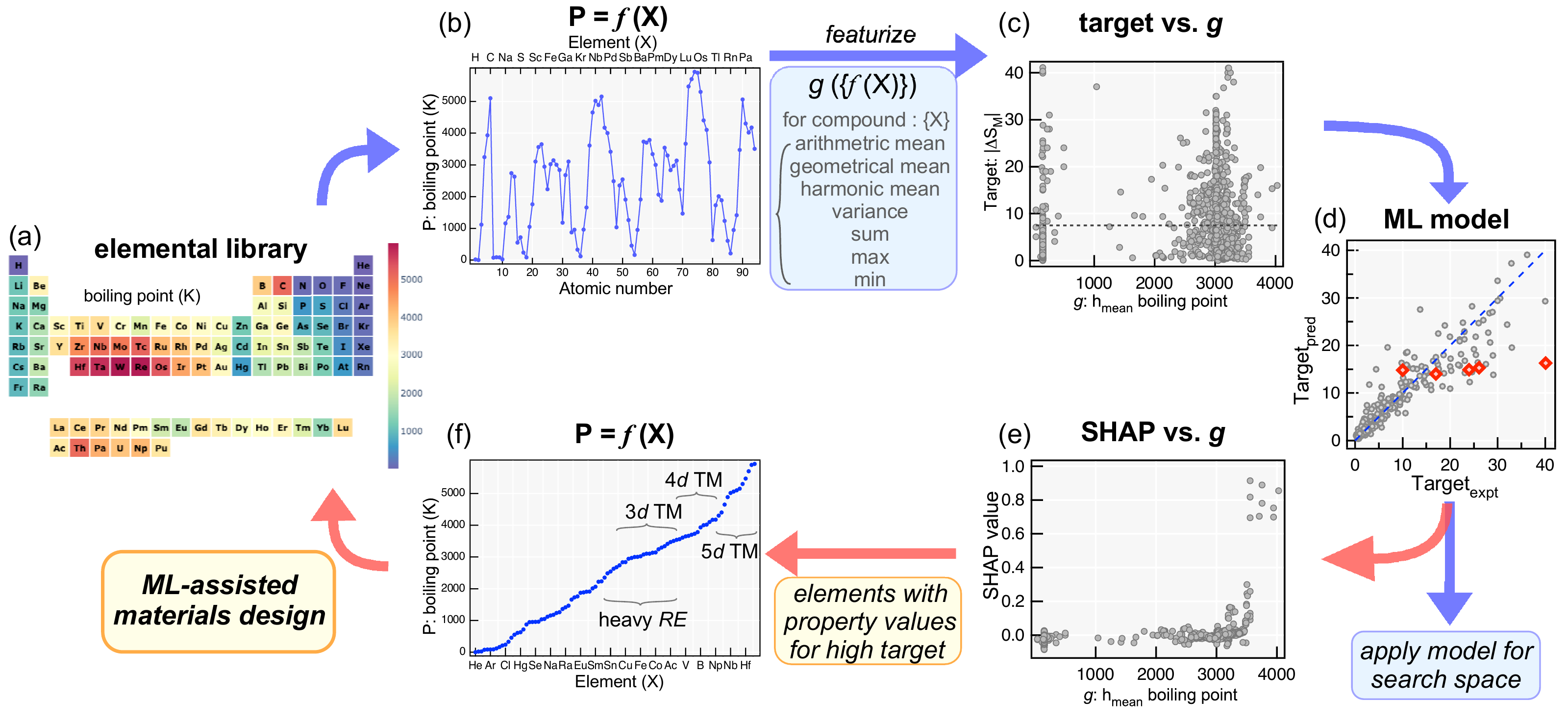}\hspace{5pt}
  \caption{Schematic workflow of materials search using compositional descriptor-based machine-learning model (blue arrows) and proposed possible materials design assisted by SHAP analysis of the model. (a)(b) Focused physical property $f(X)$ for atomic elements (X). (c) Target plotted against compositional descriptor $g(\{f(X)\})$. (d) Constructed model by regression of data in (c). (e) SHAP value plotted against compositional descriptor $g$. (f) Sorted physical property for atomic elements (X).}
  \label{fig9}
\end{figure*}

Figure 9 shows the relationship among constituent atom X, atomic element property $P$=$f(X)$, compound \{X\}, associated compositional descriptor $g(\{f(X)\})$, and target property/SHAP value of the model. As shown in Figure 9 (a)-(c), preparation of compositional descriptors corresponds to mapping each set of elements in compounds into the basis axis for the target function.  Construction of a machine-learned model corresponds to forming of the SHAP values that behave like a function against each compositional descriptor $g(\{f(X)\})$, constructed through the regression process of training data with regularization equipped in the used algorithm.  With SHAP plots like Figure 9(e), we can readily expect that when the input constituent atoms or their ratio changes it would trigger a shift in the compositional descriptor value causing a change in the output SHAP values.  It can be said that this is how the model can address the target with composition-dependence from a viewpoint of SHAP analysis.

By considering SHAP values as functions for descriptors, we can further exploit the understanding of the model. Figure 9 also shows a schematic of the conventional workflow of data-driven search using a machine-learned model with compositional descriptors (blue arrows), and an additional process that can be used for materials design and extraction of knowledge from the built model (red arrows) proposed here. In a conventional workflow, a dataset is prepared first, that contains compositions of materials consisting of a set of elements \{X\} and target physical property values. Then, compositional descriptors are generated by combining the physical properties of atomic elements with diverse mathematical operations as previously described (Figure 9(a)-(c)), that are used as training data. Once the model is built (Figure 9(d)) by using a certain algorithm, it can be used to predict target values of unknown compounds after the preparation of corresponding descriptors.  In this process, the prediction is solely relying on the model output and researchers have no access to the trend found by the model during its construction.  On the other hand, one can also perform SHAP analysis (Figure 9(e)) on the model to visualize what descriptor values tend to have high coalitions to the target value.  In the case of compositional descriptors, it can be resolved which atomic element is favorable for high SHAP as well as target values, by comparing these SHAP plots with elemental physical properties (Figure 9(f)). For instance in Figure 9, compounds with harmonic mean of boiling point of constituent atom higher than $\sim$3500 K tend to have high SHAP values, hence transition metals can be suitable as a partner element for heavy RE atoms.  We note here that changing a constituent atom to modify the value of a specific compositional descriptor can result in a change in other compositional descriptors at the same time, thus one has to be careful during such an approach.  Therefore this approach would be more effective when the SHAP values are condensed in a small number of descriptors, or when the number of prepared descriptors is reduced by pre-processing before the construction of the final model. During such pre-processing, one has to be careful since it is important to build an model as accurate as possible before applying SHAP analysis to extract plausible knowledge out of the available dataset.

\begin{figure*}[!htpb]
  \centering
  \includegraphics[width=14.5 cm]{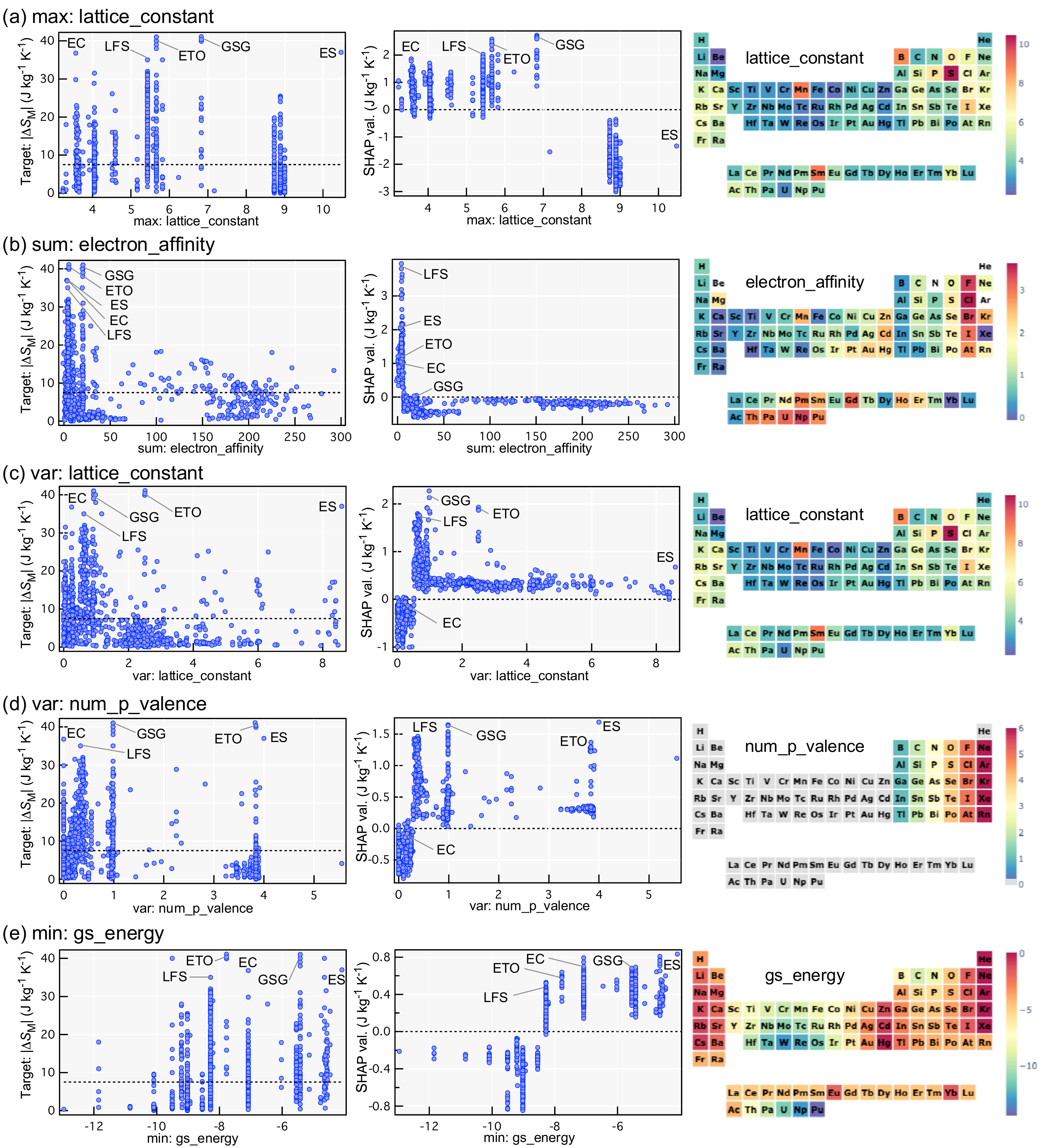}\hspace{5pt}
  \caption{Left panel: Training data distribution of target plotted against focused compositional descriptors.  Middle panel: The same as the left panel for SHAP values. Right panel: Heatmap of properties of atomic elements stored in XenonPy \cite{XenonPy}. Focused compositional descriptors are the ones that exhibited the top 5 largest absolute mean SHAP values in the model (shown in Figure 6(a)), namely (a) maximum of lattice constant, (b) sum of electron affinity, (c) variance of lattice constant, (d) variance of the number of valence $p$-electrons, and (e) minimum of ground state energy in first principles software VASP \cite{VASP}. Datapoints for several representative magnetocaloric materials are indicated explicitly, Gd$_5$Si$_{1.5}$Ge$_{2.5}$ \cite{Pecharsky_2003}, EuTiO$_3$ \cite{EuTiO3}, EuS \cite{EuS}, ErCo$_2$ \cite{Wada}, La$_{0.8}$Ce$_{0.2}$Fe$_{11.7}$Si$_{1.3}$ \cite{LaFeSi} and abbreviated as GSG, ETO, ES, EC, LFS, respectively.}
  \label{fig10}
\end{figure*}

Finally, we show further a few examples of how the constructed model with compositional descriptors and its SHAP analysis would give us an overall idea to design materials holding high target properties.  Figure 10 shows the distribution of the training dataset and associated physical properties of atomic elements (right panel), for the top 5 compositional descriptors holding high averaged SHAP values shown in Figure 6(a). That are, (a) maximum of lattice constant (\AA), (b) sum of electron affinity (eV), (c) variance of lattice constant (\AA), (d) variance of the number of valence $\textit {p}$-electrons, (e) ground state energy per atom (eV) in first principles calculation software VASP \cite{VASP}. The left panel shows the target plotted against compositional descriptors while the middle panel shows SHAP values for the same. In figures, we also highlight well-known magnetocaloric materials \cite{Franco2018} that were in the training data for a field change of 5 T, namely Gd$_5$Si$_{1.5}$Ge$_{2.5}$ (GSG) \cite{Pecharsky_2003}, EuTiO$_3$ (ETO) \cite{EuTiO3}, EuS (ES) \cite{EuS}, ErCo$_2$ (EC) \cite{Wada}, La$_{0.8}$Ce$_{0.2}$Fe$_{11.7}$Si$_{1.3}$ (LFS) \cite{LaFeSi}, whose $|\Delta S_M|$ are 41.0, 40.4, 37.0, 36.8, 34.9 (J kg$^{-1}$ K$^{-1}$), respectively.  Interestingly, those representative materials tend to lie in the realm of positive high SHAP values in the middle panel, indicating that the regression process of the model is working fine.  It is also worth noting that SHAP values for those 5 compositional descriptors tend to hold clear thresholds for positive and negative values, possibly because the model is based on a tree-based algorithm.

By combining the SHAP analysis shown in the middle panel with the property value of atomic elements shown in the right panel, we can extract several suggestions from the current model. For instance,
\begin{itemize}
 \item From Figure 10(a), the model suggests that the inclusion of atomic elements with a ``lattice constant" larger than $\sim$7 \AA $ $ tends to lower the predicted values.  Therefore the current model implies to avoid using B, S, Mn, and Sm.
 \item From Figure 10(b), the model suggests that inclusion of more than 4 or 5 atomic species would not do good for obtaining high $|\Delta S_M|$. This might make sense as such partially-substituted alloys and composites tend to be used for obtaining table-like $|\Delta S_M|$ as a function of temperature \cite{Pedro_Gd, Takeya_1994, Li_2017}.
 \item From Figure 10(c), finite variance in lattice constants of constituent atoms tends to enhance the target. Thus, a combination of anion and cation atoms with close values of elemental lattice constant can be avoided.
 \item From Figure 10(d), finite variance in the number of \textit {p}-valence electrons tends to enhance the target. Therefore, the use of anion atoms with \textit {p}-valence electrons is preferred. 
 \item From Figure 10(e), compositions containing atoms with ground state energy lower than -8.5 eV tend to have lower target values.  Thus the use of 5\textit {d}-transition metal can be avoided.
\end{itemize}

We stress here that those above are merely an example of what kind of ideas materials researchers can receive through such analysis, and the current model is yet far from complete as it is evident in the quantitative disagreement in the magnitude of the target value in the REB$_2$ system.  However, at least even in the framework of the current model, if we could find a compound satisfying those above 5 suggestions from the model, the expected target value from the sum of those 5 SHAP values of compositional features can be as high as 15-20 (J kg$^{-1}$ K$^{-1}$), which is appreciatively large, though SHAP values of rest 403 descriptors would modify the final predicted target value. We also stress that the current model holds 408 descriptors and SHAP values are widely split into small values for each descriptor.  Therefore, building a model with a confined number of descriptors and/or applying feature selection would benefit the extraction of ideas from the model through such analysis.

\section{Summary}

In summary, we have investigated the magnetocaloric effect in ErB$_2$, which was the last unreported magnetocaloric material in REB$_2$ ferromagnets.  The experimentally observed magnetic entropy change $|\Delta S_M|^{peak}$ were as large as 26.1 (J kg$^{-1}$ K$^{-1}$) for field change of 5 T at $\sim$14 K, hence $|\Delta S_M|^{peak}$ in HoB$_2$ (40.1 J kg$^{-1}$ K$^{-1}$) turned out to be the highest among REB$_2$ series (RE = Tb, Dy, Ho, Er, Tm).  The machine-learned model also predicted HoB$_2$ as the most promising material, however, the predicted magnitude of $|\Delta S_M|^{peak}$ showed a disagreement with experimental values in their quantity.  Through SHAP analysis of the model, it turned out that RE-dependence in the model prediction comes from the sum of small contributions from a number of compositional descriptors with a variety of atomic-species dependence in the current model, that might have caused a systematic error in the predicted values in REB$_2$ system. It has been also discussed that such SHAP analysis for the compositional descriptor-based machine-learned model could be helpful for researchers to visualize what trend in the training data has been found by the built model, and to plan the material design by taking advantage of this knowledge. 

\section*{Acknowledgements}

This work was supported by the JST-Mirai Program (Grant No. JPMJMI18A3), JSPS Bilateral Program (JPJSBP120214602), and JSPS KAKENHI (Grant Nos. 20K05070, 23K04572, 19H02177). P.B.C. acknowledges the scholarship support from the Ministry of Education, Culture, Sports, Science and Technology (MEXT), Japan.

\section*{Disclosure statement}

The authors declare no conflict of interest.


\bibliography{references}

\end{document}